\documentclass[preprint,eqsecnum,aps,nofootinbib]{revtex4}
\usepackage{amsfonts,amsmath,amssymb,amsthm}
\usepackage{latexsym}
\usepackage{bbm,bm}
\usepackage{graphicx}

\newcommand{\beq}{\begin{equation}}
\newcommand{\eeq}{\end{equation}}
\newcommand{\beqs}{\begin{eqnarray}}
\newcommand{\eeqs}{\end{eqnarray}}

\begin{document}

\title{R\'{e}nyi and von Neumann entropies for various Bipartite Gaussian States}

\author{DaeKil Park$^{1,2}$\footnote{dkpark@kyungnam.ac.kr}}

\affiliation{$^1$Department of Electronic Engineering, Kyungnam University, Changwon
                 631-701, Korea    \\
             $^2$Department of Physics, Kyungnam University, Changwon
                  631-701, Korea    
                      }

\begin{abstract}
The R\'{e}nyi and von Neumann entropies of various bipartite Gaussian states are derived analytically. We also discuss on the tripartite purification
for the bipartite states when some particular conditions hold. The generalization to non-Gaussian states is briefly discussed. 

\end{abstract}

\maketitle

\section{Introduction}
As IC (integrated circuit) becomes smaller and smaller, the effect of quantum mechanics becomes prominent more and more. As a result, quantum technology (technology 
based on quantum information theories\cite{text}) becomes important more and more recently. The representative constructed by quantum technology
is a quantum computer\cite{supremacy-1}, which was realized recently by making use of superconducting qubits. 

In quantum information processing entanglement\cite{schrodinger-35,horodecki09} plays an important role as a physical resource. 
It is used in various quantum information processing, such as  quantum teleportation\cite{teleportation,Luo2019},
superdense coding\cite{superdense}, quantum cloning\cite{clon}, quantum cryptography\cite{cryptography,cryptography2}, quantum
metrology\cite{metro17}, and quantum computer\cite{supremacy-1,qcreview,computer}. Furthermore, with many researchers trying to realize such quantum information processing in the laboratory for the last few decades, quantum cryptography and quantum computer seem to approaching the commercial level\cite{white,ibm}.

For last few decades many entanglement measures have been developed. For example, the entanglement of formation (EoF)\cite{benn96}, one of the
measure for bipartite system, is defined as a
von Neumann entropy of the substate in bipartite quantum state. In two qubit system the EoF of any pure and mixed states can be computed if one follows
the Wootters procedure\cite{form2,form3}. However, computation of EoF in qudit or continuous-variable (CV) system is extremely difficult problem 
except few rare cases. In this reason recent quantum technology is developed by making use of qubits generated through various physical setup.

Although quantum information theories have been developed mainly with qubit or qudit system, entanglement of CV states was also studied for last few
years. In qubit and qudit systems the necessary and sufficient criterion of separability for given quantum state is a positive partial transposition (PPT)\cite{peres96,horodecki96,horodecki97}, which is valid  for only $2 \times 2$ qubit-qubit and $2 \times 3$ qubit-qudit states.
However, as Ref. \cite{duan-2000,simon-2000} have shown, PPT also provides 
a necessary and sufficient criterion for the separability of Gaussian CV quantum states. 
Thus, the entanglement measure like negativity\cite{vidal01} can be used as a quantification of entanglement in the Gaussian states. 
Furthermore, the distillation protocols for partially entangled Gaussian state to a maximally entangled state have already been suggested 
in Duan et al.\cite{duan-99-p} and Giedke et al.\cite{giedke-2000}. Quantum information processing with CV was reviewed from both theoretical and experimental aspects\cite{braunstein-2005,adesso-2014}.

The analytical expression of von Neumann entropy was derived for a general real Gaussian density matrix in 
Ref. \cite{luca86} and it was generalized to massless scalar field in Ref. \cite{mark93}. Putting the scalar field system in the spherical box, the author 
in Ref. \cite{mark93} has shown that the total entropy of the system is proportional to surface area. This may give some insight into the reason why the black hole 
entropy is proportional to the horizon area. Recently, the entanglement was computed in the coupled harmonic oscillator  system using a Schmidt decomposition\cite{maka17}. The von Neumann and R\'{e}nyi entropies were also explicitly computed in the similar system, called two site Bose-Hubbard model\cite{ghosh17}. More recently, the R\'{e}nyi and von Neumann entropies in bipartite\cite{park18-1}, tripartite\cite{park18-2}, and 
thermal\cite{park19} harmonic oscillator systems were computed when the spring and coupling constants are arbitrarily time-dependent. 

In this paper we will compute  the R\'{e}nyi and von Neumann entropies of various bipartite Gaussian states, which can be substates of muitipartite 
Gaussian states or their partial transpose states. Thus, the bipartite states we consider in this paper can be physically relevant or irrelevant. We categorize
the bipartite Gaussian states into four types and compute their  R\'{e}nyi and von Neumann entropies. Probably, many bipartite Gaussian states
belong to one of four types. 

The paper is organized as follows. In section II we compute  the R\'{e}nyi and von Neumann entropies for the most general single-party 
Gaussian states. The results of this section are used in next section if the eigenvalue equation of the bipartite density matrix can 
be factorized into two single-party eigenvalue equations. In section III we compute  the R\'{e}nyi and von Neumann entropies of four types of 
bipartite Gaussian states analytically. For type I, II, and III the entropies are exactly derived. For the case of type IV we cannot find a 
suitable change of variables, which factorizes the eigenvalue equation into two single-party eigenvalue equations. We compute the entropies of 
type IV under some assumption (see Eq. (\ref{m-eigenvalue-2})).  In Sec. IV a brief conclusion is given.

\section{The Most general single-party Gaussian State}
Before we consider various bipartite Gaussian states in next section, we derive the entropies and spectral decomposition of the most general single-party 
Gaussian quantum state given by 
\begin{equation}
\label{single-p}
\rho_0 [x', x] = A e^{-a_1 x^2 - a_2 x'^2 + 2 b x x'}
\end{equation}
where  $A = \sqrt{\frac{a_1 + a_2 - 2 b}{\pi}}$. In order to compute the R\'{e}nyi and von Neumann entropies for $\rho_0$ we should solve 
the following eigenvalue equation:
\begin{equation}
\label{eigen2-1}
\int \rho_0 [x',x] f_n (x) dx = \lambda_n f_n (x')
\end{equation}
with $n = 0, 1, 2, \cdots$. One can show directly that the eigenfunction is 
\begin{equation}
\label{eigen2-2}
f_n (x) = {\cal C}_n^{-1} H_n (\sqrt{\epsilon_0} x) e^{-\frac{\alpha_0}{2} x^2}
\end{equation}
where $\epsilon_0 = \sqrt{(a_1 + a_2)^2 - 4 b^2}$ and $\alpha_0 = \epsilon_0 - (a_1 - a_2)$, and $H_n (z)$ is $n^{th}$-order Hermite polynomial. 
By making use of integral formula\cite{integral}
\begin{eqnarray}
\label{integral2-1}
&&\int_{-\infty}^{\infty} e^{-(x - y)^2} H_{m} (c x) H_n (c x) d x      \\     \nonumber
&&= \sqrt{\pi} \sum_{k=0}^{\min (m, n)} 2^k k! \left(  \begin{array}{c} m  \\  k  \end{array}  \right) 
\left(  \begin{array}{c} n  \\  k  \end{array}  \right) (1 - c^2)^{\frac{m+n}{2} - k} H_{m+n - 2 k} \left( \frac{c y}{\sqrt{1 - c^2}} \right)
\end{eqnarray}
and various properties of Gamma function\cite{handbook}, the normalization constant ${\cal C}_n$ can be written in a form 
\begin{equation}
\label{normalization2-1}
{\cal C}_n^2 = \frac{1}{\sqrt{\alpha_0}} \sum_{k=0}^n 2^{2 n - k} \left( \frac{\epsilon_0}{\alpha_0} - 1 \right)^{n-k}
\frac{\Gamma^2 (n+1) \Gamma \left(n - k + \frac{1}{2} \right)}{\Gamma (k+1) \Gamma^2 (n - k + 1)}.
\end{equation}
If $a_1 = a_2$, $\alpha_0 = \epsilon_0$, which makes nonzero in $k$-summation of Eq. (\ref{normalization2-1}) only when $k = n$. Then, 
${\cal C}_n$ becomes usual harmonic oscillator normalization constant 
\begin{equation}
\label{normalization2-2}
{\cal C}_n ^{-1} = \frac{1}{\sqrt{2^n n!}} \left( \frac{\epsilon_0}{\pi} \right)^{1 / 4}.
\end{equation}

It is straightforward to show that the eigenvalues in Eq. (\ref{eigen2-1}) become
\begin{equation}
\label{eigen2-3}
\lambda_n = A \sqrt{\frac{2 \pi}{(a_1 + a_2) + \epsilon_0}} \left[ \frac{(a_1 + a_2) - \epsilon_0}{(a_1 + a_2) + \epsilon_0} \right]^{n/2} =  (1 - \xi_0) \xi_0^n
\end{equation}
where 
\begin{equation}
\label{eigen2-4}
\xi_0 = \frac{2 b}{a_1 + a_2 + \epsilon_0} = \frac{\sqrt{a_1 + a_2 + 2 b} - \sqrt{a_1 + a_2 - 2 b}}{\sqrt{a_1 + a_2 + 2 b} + \sqrt{a_1 + a_2 - 2 b}}.
\end{equation}
Eq. (\ref{eigen2-3}) is consistent with $\mbox{tr} \rho_0 = 1$ if $|\xi_0| < 1$. Thus, $\rho_0$ is physical quantum state if  $0 < \xi_0 < 1$. 
If this condition does not hold, $\rho_0$ is unphysical Gaussian state. Anyway, the R\'{e}nyi and von Neumann entropies for $\rho_0$ are 
\begin{equation}
\label{entropy2-1}
S_{0, \alpha} = \frac{1}{1 - \alpha} \ln \frac{(1 - \xi_0)^{\alpha}}{1 - \xi_0^{\alpha}}    \hspace{1.0cm}
S_{0,von} = -\ln (1 - \xi_0) - \frac{\xi_0}{1 - \xi_0} \ln \xi_0.
\end{equation}
Finally, the spectral decomposition of $\rho_0$ can be written as 
\begin{equation}
\label{spectral2-1}
\rho_0 [x',x] = \sum_n \lambda_n f_n (x') f_n^* (x).
\end{equation}

\section{Several Bipartite Gaussian States}
In this section we consider four types of bipartite Gaussian states $\rho_{j} [x_1', x_2': x_1, x_2]$ with $j$ = I, II, III, IV. They can be  physically relevant or 
irrelevant depending on their eigenvalues.  In order to compute 
the R\'{e}nyi and von Neumann entropies of $\rho_{j}$, we should solve the following eigenvalue equation 
\begin{equation}
\label{m-eigenvalue-1}
\int dx_1 dx_2 \rho_{j} [x_1', x_2': x_1, x_2] f_{mn} (x_1, x_2) = \lambda_{mn} f_{mn} (x_1', x_2')
\end{equation}
with $m,n = 0, 1, 2, \cdots$. Since $\mbox{tr} \rho_{j} = 1$, the eigenvalue $\lambda_{mn}$ obeys $\sum_{m,n} \lambda_{mn} = 1$. 
In this reason the eigenvalues of $\rho_{j}$ can be represented as 
\begin{equation}
\label{m-eigenvalue-2}
\lambda_{mn} = (1 - \xi_1) \xi_1^m (1 - \xi_2) \xi_2^n,
\end{equation}
where $0 < |\xi_1|, |\xi_2| < 1$. If $0 < \xi_1, \xi_2 < 1$, the bipartite Gaussian states are physically relevant. If not, they are unphysical states. 
Then, the R\'{e}nyi and von Neumann entropies of $\rho_{j}$ become
\begin{equation}
\label{m-entropy-1}
S_{\alpha} \equiv \frac{1}{1 - \alpha} \ln \mbox{tr} \rho_j^{\alpha} = S_{1,\alpha} + S_{2, \alpha}   \hspace{1.0cm} 
S_{von} \equiv \lim_{\alpha \rightarrow 1} S_{\alpha} = S_{1,von} + S_{2,von}
\end{equation}
where
\begin{equation}
\label{m-entropy-2}
S_{k, \alpha} = \frac{1}{1 - \alpha} \ln \frac{(1 - \xi_k)^{\alpha}}{1 - \xi_k^{\alpha}}       \hspace{1.0cm}
S_{k, von} = - \ln (1 - \xi_k) - \frac{\xi_k}{1 - \xi_k} \ln \xi_k 
\end{equation}
with $k= 1, 2$. Thus, in the following we will derive $\xi_1$ and $\xi_2$ for each $\rho_j$ analytically.

\subsection{Type I Bipartite State}

In this subsection we examine the following bipartite Gaussian state:
\begin{eqnarray}
\label{type1}
&& \rho_I [x_1', x_2': x_1, x_2]     
 = A \exp \bigg[ -a_1 (x_1'^2 + x_1^2) - a_2 (x_2'^2 + x_2^2) + 2 b (x_1' x_2' + x_1 x_2)      \\    \nonumber
&&  \hspace{7.0cm} + 2 c (x_1 x_1' + x_2 x_2') + 2 f (x_1 x_2' + x_2 x_1')  \bigg]
\end{eqnarray}
where $A = 2 \sqrt{(a_1 - c) (a_2 - c) - (b + f)^2} / \pi$. We will call this type of Gaussian states by type I. 
The purity function of $\rho_I$ is 
\begin{equation}
\label{purity-1}
\mbox{tr} \rho_I^2 = \sqrt{\frac{(a_1 - c) (a_2 - c) - (b + f)^2}{(a_1 + c) (a_2 + c) - (b - f)^2}}.
\end{equation}
Thus, if $c = f = 0$, $\rho_I$ becomes pure state. For other cases it is mixed state. 

Now, let us solve the eigenvalue equation (\ref{m-eigenvalue-1}) (with $j$ = I) explicitly. 
First, we change the variables as 
\begin{eqnarray}
\label{change3-1}
\left(    \begin{array}{c} X_1  \\  X_2    \end{array}    \right)
=    \left(     \begin{array}{cc}
                 \cos \theta  &   -\sin \theta   \\
                 \sin \theta   &    \cos \theta   
                    \end{array}                          \right)
\left(    \begin{array}{c} x_1  \\  x_2    \end{array}    \right)
\end{eqnarray}
where 
\begin{equation}
\label{angle3-1}
\theta = \tan^{-1} \frac{2 b } {\sqrt{(a_1 - a_2)^2 + 4 b^2} + (a _1 - a_2)}.
\end{equation}
In terms of new variables the eigenvalue equation (\ref{m-eigenvalue-1}) can be written as 
\begin{eqnarray}
\label{eigen3-2}
&&A e^{-\mu_+ X_1'^2 - \mu_- X_2'^2} \int d X_1 dX_2 e^{-\mu_+ X_1^2 - \mu_- X_2^2 + 2 C_1 X_1 X_1' + 2 C_2 X_2 X_2' + 2 F (X_1 X_2' + X_2 X_1')}
f_{mn} (X_1, X_2 )                                                                                      \nonumber     \\
&&   \hspace{7.0cm}  = \lambda_{mn} f_{mn} (X_1', X_2')
\end{eqnarray}
where
\begin{eqnarray}
\label{eigen3-3}
&&\hspace{1.5cm}\mu_{\pm} = \frac{1}{2} \left[ (a_1 + a_2) \pm \sqrt{(a_1- a_2)^2 + 4 b^2} \right]      \\     \nonumber
&&C_1 = c - f \sin 2\theta   \hspace{1.0cm}  C_2 = c + f \sin 2 \theta   \hspace{1.0cm}  F = f \cos 2 \theta.
\end{eqnarray}

In order to simplify Eq. (\ref{eigen3-2}) some more we define 
\begin{equation}
\label{change3-2}
y_1 = \sqrt{\mu_+} X_1   \hspace{2.0cm}  y_2 = \sqrt{\mu_-} X_2.
\end{equation}
Then, Eq. (\ref{eigen3-2}) reduces to 
\begin{eqnarray}
\label{eigen3-4}
&&\frac{A}{\sqrt{\mu_+ \mu_-}} e^{-y_1'^2 - y_2'^2} 
\int dy_1 dy_2 \exp \left[-y_1^2 - y_2^2 + \frac{2 C_1}{\mu_+} y_1 y_1' + \frac{2C_2}{\mu_-} y_2 y_2' + \frac{2 F}{\sqrt{\mu_+ \mu_-}}  
(y_1 y_2' + y_2 y_1')  \right]                                                                                    \nonumber \\
&& \hspace{5.0cm} \times f_{mn} (y_1, y_2) = \lambda_{mn} f_{mn} (y_1', y_2').
\end{eqnarray}

Finally, we change the variables again as 
\begin{eqnarray}
\label{change3-3}
\left(    \begin{array}{c} Y_1  \\  Y_2    \end{array}    \right)
=    \left(     \begin{array}{cc}
                 \cos \varphi  &   -\sin \varphi   \\
                 \sin \varphi   &    \cos \varphi   
                    \end{array}                          \right)
\left(    \begin{array}{c} y_1  \\  y_2    \end{array}    \right)
\end{eqnarray}
where 
\begin{equation}
\label{angle3-2}
\varphi = \tan^{-1} \left[ \frac{-2 F / \sqrt{\mu_+ \mu_-}}
{\sqrt{\left( \frac{C_1}{\mu_+} - \frac{C_2}{\mu_-} \right)^2 + \frac{4 F^2}{\mu_+ \mu_-}} + \left( \frac{C_1}{\mu_+} - \frac{C_2}{\mu_-} \right)}  \right].
\end{equation}
In terms of the new coordinates Eq. (\ref{eigen3-4}) is simplified in a form 
\begin{equation}
\label{eigen3-5}
\frac{A}{\sqrt{\mu_+ \mu_-}} e^{-Y_1'^2 - Y_2'^2} \int dY_1 dY_2 
e^{-Y_1^2 - Y_2^2 + 2 \nu_+ Y_1 Y_1' + 2 \nu_- Y_2 Y_2'} f_{mn} (Y_1, Y_2) = \lambda_{mn} f_{mn} (Y_1', Y_2')
\end{equation}
where 
\begin{equation}
\label{eigen3-6}
\nu_{\pm} = \frac{1}{2} \left[ \left( \frac{C_1}{\mu_+} + \frac{C_2}{\mu_-} \right) \pm 
\sqrt{\left( \frac{C_1}{\mu_+} - \frac{C_2}{\mu_-} \right)^2 + \frac{4 F^2}{\mu_+ \mu_-}} \right].
\end{equation}

Now, we define 
\begin{equation}
\label{define3-1}
f_{mn} (Y_1, Y_2) = G_{1,m} (Y_1) G_{2,n} (Y_2).
\end{equation}
Then, the eigenvalue equation (\ref{eigen3-5}) is solved if one solves the following two single-party eigenvalue equations:
\begin{eqnarray}
\label{eigen3-7}
&&e^{-Y_1'^2} \int dY_1 e^{-Y_1^2 + 2 \nu_+ Y_1 Y_1'} G_{1,m} (Y_1) = p_m G_{1,m} (Y_1')      \\    \nonumber
&&e^{-Y_2'^2} \int dY_2 e^{-Y_2^2 + 2 \nu_- Y_2 Y_2'} G_{2,n} (Y_2) = q_n G_{2,n} (Y_2').
\end{eqnarray}
Using the results of the previous section it is easy to show that the eigenvalues $p_m$ and $q_n$ are 
\begin{equation}
\label{eigen3-8}
p_m = \sqrt{\frac{\pi}{1 + \epsilon_+}} \left( \frac{1 - \epsilon_+}{1 + \epsilon_+} \right)^{m/2}       \hspace{1.0cm}
q_n = \sqrt{\frac{\pi}{1 + \epsilon_-}} \left( \frac{1 - \epsilon_-}{1 + \epsilon_-} \right)^{n/2}  
\end{equation}    
where $\epsilon_{\pm} = \sqrt{1 - \nu_{\pm}^2}$. Then, the eigenvalue $\lambda_{mn} = A p_m q_n / \sqrt{\mu_+ \mu_-}$ 
can be represented as Eq. (\ref{m-eigenvalue-2}), where 
\begin{equation}
\label{eigen3-10}
\xi_1 = \frac{\nu_+}{1 + \epsilon_+} = \frac{\sqrt{1 + \nu_+} - \sqrt{1 - \nu_+}}{\sqrt{1 + \nu_+} + \sqrt{1 - \nu_+}}   \hspace{1.0cm}
\xi_2 = \frac{\nu_-}{1 + \epsilon_-} = \frac{\sqrt{1 + \nu_-} - \sqrt{1 - \nu_-}}{\sqrt{1 + \nu_-} + \sqrt{1 - \nu_-}}. 
\end{equation}
Thus, the R\'{e}nyi and von Neumann entropies of $\rho_I$ can be explicitly computed by inserting $\xi_1$ and $\xi_2$ in Eq. (\ref{eigen3-10}) 
into Eq. (\ref{m-entropy-1}) and (\ref{m-entropy-2}). 
Applying the results of the previous section one can derive the normalized eigenfunction, whose explicit expression is 
\begin{equation}
\label{eigen3-13}
f_{mn} (x_1, x_2) = \left[ \frac{1}{\sqrt{2^m m!}} \left(\frac{2 \epsilon_+}{\pi} \right)^{1/4} H_m (\sqrt{2 \epsilon_+} Y_1) e^{-\epsilon_+ Y_1^2} \right]
\left[ \frac{1}{\sqrt{2^n n!}} \left(\frac{2 \epsilon_-}{\pi} \right)^{1/4} H_n (\sqrt{2 \epsilon_-} Y_2) e^{-\epsilon_- Y_2^2} \right].
\end{equation}
Thus, the spectral decomposition of $\rho_I$ is 
\begin{equation}
\label{sepctral3-1}
\rho_I [x_1', x_2':x_1, x_2] = \sum_{m,n} \lambda_{mn} f_{mn} (x_1', x_2') f_{mn}^* (x_1, x_2).
\end{equation}

If $c = \pm f$, either $\xi_1$ or $\xi_2$ is zero depending on the signs of $c$, $f$, and $(\mu_+ + \mu_-) \pm (\mu_+ - \mu_-) \sin 2 \theta$. 
This fact implies from Schmidt decomposition that $\rho_I$ can be purified to tripartite pure state. If, for example, $c = f  \equiv z$, the tripartite pure state 
can be constructed as 
\begin{equation}
\label{puri3-1} 
\sigma_{ABC} [x_1', x_2', x_3': x_1, x_2, x_3] = \psi(x_1',x_2',x_3') \psi^* (x_1,x_2, x_3)
\end{equation}
where 
\begin{eqnarray}
\label{puri3-2}
&&\psi (x_1,x_2, x_3) = \frac{1}{\pi^{3/4}} \exp \bigg[ -(a_1 + z) x_1^2 - (a_2 + z) x_2^2 - \frac{x_3^3}{8 [(a_1 - z) (a_2 - z) - (b+z)^2]}   \nonumber \\
&&  \hspace{3.5cm}+ 2 (b - z) x_1 x_2 - \sqrt{\frac{z}{(a_1 - z) (a_2 - z) - (b+z)^2}} (x_1 + x_2) x_3 \bigg].
\end{eqnarray}
Then, $\rho_I$ is constructed by $\rho_I = \mbox{tr}_C \sigma_{ABC}$. If $c = -f \equiv z$, $\psi(x_1,x_2, x_3)$ in Eq. (\ref{puri3-2}) is 
changed into 
\begin{eqnarray}
\label{puri3-3}
&&\psi (x_1,x_2, x_3) = \frac{1}{\pi^{3/4}} \exp \bigg[ -(a_1 + z) x_1^2 - (a_2 + z) x_2^2 - \frac{x_3^3}{8 [(a_1 - z) (a_2 - z) - (b-z)^2]}   \nonumber \\
&&  \hspace{3.5cm}+ 2 (b + z) x_1 x_2 - \sqrt{\frac{z}{(a_1 - z) (a_2 - z) - (b-z)^2}} (x_1 - x_2) x_3 \bigg].
\end{eqnarray}

\subsection{Type II Bipartite State}

In this subsection we examine the following bipartite Gaussian state:
\begin{eqnarray}
\label{type2}
&& \rho_{II} [x_1', x_2': x_1, x_2]     
 = A \exp \bigg[ -a_1 (x_1'^2 + x_2'^2) - a_2 (x_1^2 + x_2^2) + 2 b_1 x_1' x_2' + 2 b_2 x_1 x_2     \\    \nonumber
&&  \hspace{7.0cm} + 2 c (x_1 x_1' + x_2 x_2') + 2 f (x_1 x_2' + x_2 x_1')  \bigg]
\end{eqnarray}
where $A =  \sqrt{(a_1 + a_2 - 2c)^2 - (b_1 + b_2 + 2 f)^2} / \pi$. We will call this type of Gaussian states by type II. This type of Gaussian state does not 
arise in the usual harmonic oscillator systems. This means that this type of states cannot be obtained as a substate of mutipartite pure state in the usual harmonic oscillator system. However, this type appears in Ref.\cite{park19} as a thermal state in the two-coupled harmonic oscillator system.
The purity function of $\rho_{II}$ is 
\begin{equation}
\label{purity-2}
\mbox{tr} \rho_{II}^2 = \sqrt{\frac{(a_1 + a_2 - 2 c)^2 - (b_1 + b_2 + 2 f)^2}{(a_1 + a_2 + 2 c)^2 - (b_1 + b_2 - 2 f)^2}}.
\end{equation}
Thus, if $c = f = 0$, $\rho_{II}$ is pure Gaussian state. If not, it is mixed state. 

Now, let us solve the eigenvalue equation (\ref{m-eigenvalue-1}) (with $j$ = II) analytically. 
First we change the variables as 
\begin{equation}
\label{change4-1}
y_1 = \frac{1}{\sqrt{2}} (x_1 + x_2)   \hspace{2.0cm}  y_2 = \frac{1}{\sqrt{2}} (x_1 - x_2).
\end{equation}
Then, Eq. (\ref{m-eigenvalue-1}) is simplified as 
\begin{eqnarray}
\label{eigen4-2}
&& A e^{-(a_1 - b_1) y_1'^2 - (a_1 + b_1) y_2'^2} 
\int dy_1 dy_2 e^{- (a_2 - b_2) y_1^2 - (a_2 + b_2) y_2^2 + 2 (c + f) y_1' y_1 + 2 (c - f) y_2' y_2} f_{mn} (y_1, y_2)   \nonumber   \\
&&  \hspace{8.0cm}= \lambda_{mn} f_{mn} (y_1', y_2').
\end{eqnarray}
If we define 
\begin{equation}
\label{define4-1}
f_{mn} (y_1, y_2) = g_m (y_1) h_n (y_2),
\end{equation}
Eq. (\ref{eigen4-2}) can be solved if one solves the following two single-party eigenvalue equations:
\begin{eqnarray}
\label{eigen4-3}
&& e^{-(a_1 - b_1) y_1'^2} \int dy_1 e^{-(a_2 - b_2) y_1^2 + 2 (c + f) y_1' y_1} g_m (y_1) = p_m g_m (y_1')       \\     \nonumber
&& e^{-(a_1 + b_1) y_2'^2} \int dy_2 e^{-(a_2 + b_2) y_2^2 + 2 (c - f) y_2' y_2} h_n (y_2) = q_n h_n (y_2').
\end{eqnarray}
The solutions of Eq. (\ref{eigen4-3}) are analytically derived by using the results of the previous section. 
Then, eigenvalue of Eq. (\ref{m-eigenvalue-1}) is $\lambda_{mn} = A p_m q_n$, which can be represented as Eq. (\ref{m-eigenvalue-2}), where 
\begin{eqnarray}
\label{eigen4-4}
&&\xi_1 = \frac{2 (c + f)}{(a_1 + a_2 - b_1 - b_2) + \epsilon_1}                                                                          \\   \nonumber
&&\hspace{.5cm}= \frac{\sqrt{(a_1 + a_2 - b_1 - b_2) + 2 (c+f)} - \sqrt{(a_1 + a_2 - b_1 - b_2) - 2 (c+f)}}
                                                                                                        {\sqrt{(a_1 + a_2 - b_1 - b_2) + 2 (c+f)} + \sqrt{(a_1 + a_2 - b_1 - b_2) - 2 (c+f)}}
                                                                                                                                                                        \\   \nonumber
&&\xi_2 = \frac{2 (c - f)}{(a_1 + a_2 + b_1 + b_2) + \epsilon_2}                                                                          \\   \nonumber
&&\hspace{.5cm}= \frac{\sqrt{(a_1 + a_2 + b_1 + b_2) + 2 (c-f)} - \sqrt{(a_1 + a_2 + b_1 + b_2) - 2 (c-f)}}
                                      {\sqrt{(a_1 + a_2 + b_1 + b_2) + 2 (c-f)} + \sqrt{(a_1 + a_2 + b_1 + b_2) - 2 (c-f)}}
\end{eqnarray}
with
\begin{equation}
\label{eigen4-5}
\epsilon_1 = \sqrt{(a_1 + a_2 - b_1 - b_2)^2 -4 (c + f)^2}     \hspace{1.0cm} \epsilon_2 = \sqrt{(a_1 + a_2 + b_1 + b_2)^2 -4 (c - f)^2}.
\end{equation}
Thus, the R\'{e}nyi and von Neumann entropies of $\rho_{II}$ can be explicitly computed by inserting $\xi_1$ and $\xi_2$ in Eq. (\ref{eigen4-4}) 
into Eq. (\ref{m-entropy-1}) and (\ref{m-entropy-2}). 

Following the analysis in section II one can also derive the normalized eigenfunction, whose explicit expression is 
\begin{equation}
\label{eigen4-6}
f_{mn} (x_1, x_2) = \left({\cal C}_{1,m}^{-1} H_m (\sqrt{\epsilon_1} y_1) e^{-\frac{\alpha_1}{2} y_1^2} \right) 
                            \left({\cal C}_{2,n}^{-1} H_n (\sqrt{\epsilon_2} y_2) e^{-\frac{\alpha_2}{2} y_2^2} \right)  
\end{equation}  
where 
\begin{equation}
\label{eigen4-7}
\alpha_1 = \epsilon_1 + (a_1 - a_2) - (b_1 - b_2)    \hspace{1.0cm}  \alpha_2 = \epsilon_2 + (a_1 - a_2) + (b_1 - b_2)
\end{equation}    
and the normalization constants ${\cal C}_{1,m}$ and ${\cal C}_{2,n}$ are   
\begin{eqnarray}
\label{eigen4-8}
&&{\cal C}_{1,m}^2 = \frac{1}{\sqrt{\alpha_1}} \sum_{k=0}^m 2^{2m - k} \left( \frac{\epsilon_1}{\alpha_1} - 1 \right)^{m-k} 
    \frac{\Gamma^2 (m+1) \Gamma (m - k + 1/2)}{\Gamma (k + 1) \Gamma^2 (m-k+1)}                                \\    \nonumber
&&{\cal C}_{2,n}^2 = \frac{1}{\sqrt{\alpha_2}} \sum_{k=0}^n 2^{2n - k} \left( \frac{\epsilon_2}{\alpha_2} - 1 \right)^{n-k} 
    \frac{\Gamma^2 (n+1) \Gamma (n - k + 1/2)}{\Gamma (k + 1) \Gamma^2 (n-k+1)}. 
\end{eqnarray} 
Thus, the spectral decomposition of $\rho_{II}$ is 
\begin{equation}
\label{sepctral4-1}
\rho_{II} [x_1', x_2':x_1, x_2] = \sum_{m,n} \lambda_{mn} f_{mn} (x_1', x_2') f_{mn}^* (x_1, x_2).
\end{equation}

Now, let us consider a case of $a_1 = a_2 \equiv a$ and $b_1 = b_2 \equiv b$.
In this case  $\epsilon_1$ and $\epsilon_2$ become $\epsilon_1 = 2 \sqrt{(a - b)^2 - (c + f)^2}$ and 
$\epsilon_2 = 2 \sqrt{(a + b)^2 - (c - f)^2}$. Thus, $\xi_1$, $\xi_2$, ${\cal C}_{1,m}$, and ${\cal C}_{2,n}$ become
\begin{eqnarray}
\label{special4-1}
&&\xi_1 = \frac{c + f}{(a - b) + \frac{\epsilon_1}{2}} = \frac{\sqrt{(a - b) + (c+f)} - \sqrt{(a-b) - (c+f)}}{\sqrt{(a - b) + (c+f)} + \sqrt{(a-b) - (c+f)}}
                                                                                                                                                                                           \\   \nonumber
&&\xi_2 = \frac{c - f}{(a + b) + \frac{\epsilon_2}{2}} = \frac{\sqrt{(a + b) + (c-f)} - \sqrt{(a+b) - (c-f)}}{\sqrt{(a + b) + (c-f)} + \sqrt{(a+b) - (c-f)}}
                                                                                                                                                                                            \\    \nonumber
&& {\cal C}_{1,m}^{-1} = \frac{1}{\sqrt{2^m m!}} \left(\frac{\epsilon_1}{\pi} \right)^{1/4}    \hspace{1,0cm}
 {\cal C}_{2,n}^{-1} = \frac{1}{\sqrt{2^n n!}} \left(\frac{\epsilon_2}{\pi} \right)^{1/4}.
\end{eqnarray}
Eq. (\ref{special4-1}) implies that either $\xi_1$ or $\xi_2$ is zero if $c = \pm f$, which may indicate that the tripartite purification of $\rho_{II}$ 
is possible. In fact, one can find a following tripartite pure state, whose substate coincides with $\rho_{II}$. For $c=f \equiv z$ the pure state is 
\begin{eqnarray}
\label{puri4-1}
&&\psi(x_1, x_2, x_3) = \frac{1}{\pi^{3/4}} \exp \bigg[ -(a+z) (x_1^2 + x_2^2) - \frac{x_3^2}{8 [ (a - z)^2 - (b+z)^2]}    \\   \nonumber
&&\hspace{4.0cm} + 2 (b-z) x_1 x_2 - \sqrt{\frac{z}{(a-z)^2 - (b+z)^2}} (x_1 +x_2) x_3  \bigg].
\end{eqnarray}
If $c = -f \equiv z$, the purification is changed into    
\begin{eqnarray}
\label{puri4-2}
&&\psi(x_1, x_2, x_3)= \frac{1}{\pi^{3/4}} \exp \bigg[ -(a+z) (x_1^2 + x_2^2) - \frac{x_3^2}{8 [ (a - z)^2 - (b-z)^2]}    \\   \nonumber
&&\hspace{4.0cm} + 2 (b+z) x_1 x_2 - \sqrt{\frac{z}{(a-z)^2 - (b-z)^2}} (x_1 -x_2) x_3  \bigg].
\end{eqnarray}

Even without the restriction  $a_1 = a_2 \equiv a$ and $b_1 = b_2 \equiv b$, Eq. ({\ref{eigen4-4}) implies that either $\xi_1$ or $\xi_2$ is zero at 
$c = \pm f$. Thus, one may conjecture that the tripartite purification of $\rho_{II}$ might be possible at $c = \pm f$ even in this case. However, we cannot find any pure state $\sigma_{ABC}$, which gives $\rho_{II}$ as $\rho_{II} = \mbox{tr}_C \sigma_{ABC}$. We think this is because of the fact that $\rho_{II}$ cannot arise in the usual 
zero-temperature harmonic oscillator system. 

\subsection{Type III Bipartite State}
In this subsection we examine the following bipartite Gaussian state:
\begin{eqnarray}
\label{type3-1}
&&\rho_{III} \left[ x_1', x_2': x_1, x_2 \right] = A \exp \bigg[ -a_1 (x_1'^2 + x_1^2) - a_2 (x_2'^2 + x_2^2)    \\    \nonumber
&&\hspace{1.0cm}+ 2 b (x_1 x_2 + x_1' x_2') + 2 c_1 x_1 x_1' + 2 c_2 x_2 x_2'
+ 2 f x_1' x_2 + 2 f^* x_1 x_2'  \bigg]
\end{eqnarray}
where $A = 2 \sqrt{(a_1 - c_1) (a_2 - c_2) - (b + f_R)^2} / \pi$, with $f_R = \mbox{Re} f$. We will call this type of Gaussian state by type III. 
If $c_1 = c_2 \equiv c$ and $f = f^*$, $\rho_{III}$ reduces to 
$\rho_I$. The type III Gaussian state arises in the reduced state of three coupled harmonic oscillator system\cite{park18-2}. 
In this reference the eigenvalue equation (\ref{m-eigenvalue-1}) (with $j$ = III) is analytically solved by introducing the appropriate change of variables 
as we did in the previous sections. In this paper, however, we will compute the entropy of $\rho_{III}$ without solving the eigenvalue equation explicitly as follows. This method is useful in the next subsection. 

First, we note that $\beta_1 = \mbox{tr} \rho_{III}^2$ and $\beta_2 = \mbox{tr} \rho_{III}^3$ become
\begin{equation}
\label{new5-1}
\beta_1 = \sqrt{\frac{X_1}{X_2}}                     \hspace{1.0cm}
\beta_2 = \frac{4 X_1}{X_1 + 3 X_2 - 12 x}
\end{equation}
where
\begin{eqnarray}
\label{new5-2}
&&\hspace{2.0cm}X_1 = 4 \left[ (a_1 - c_1) (a_2 - c_2) - (b + f_R)^2 \right]               \\    \nonumber
&&X_2 = 4 \left[ (a_1 + c_1) (a_2 + c_2) - (b - f_R)^2 \right]      \hspace{1.0cm}                                        
 x = c_1 c_2 - |f|^2.
\end{eqnarray}
It is easy to show $X_2 \pm X_1 = 8 A_{\pm}$ and $X_1 X_2 = 16 \tilde{A}$, where 
\begin{eqnarray}
\label{new5-3}
&&\hspace{2.0cm}A_+ = a_1 a_2 + c_1 c_2 - b^2 - f_R^2    \hspace{1.0cm}  A_- = a_1 c_2 + a_2 c_1 + 2 b f_R                \\    \nonumber
&&\tilde{A} =(a_1^2 - c_1^2) (a_2^2 - c_2^2) + (b^2 - f_R^2)^2 - 2 (a_1 a_2 + c_1 c_2) (b^2 + f_R^2) - 4 b f_R (a_1 c_2 + a_2 c_1).
\end{eqnarray}

The eigenvalue of Eq. (\ref{m-eigenvalue-1}) is shown to be represented as Eq. (\ref{m-eigenvalue-2}) in Ref. \cite{park18-2}. 
Thus, the remaining task we should do is to compute $\xi_1$ and $\xi_2$ explicitly.
Putting $\alpha=2$ or $3$ in $S_{\alpha}$ of Eq. (\ref{m-entropy-1}), it is possible to derive the following equations:
\begin{equation}
\label{negat-19}
\frac{(1 - \xi_1) (1 - \xi_2)}{(1 + \xi_1) (1 + \xi_2)} = \beta_1    \hspace{1.0cm}
\frac{(1 - \xi_1)^2 (1 - \xi_2)^2}{(1 + \xi_1 + \xi_1^2) (1 + \xi_2 + \xi_2^2)} = \beta_2.
\end{equation}
Solving Eq. (\ref{negat-19}), we get 
\begin{eqnarray}
\label{negat-22}
&&u = \xi_1 + \xi_2 = \frac{1 - \beta_1}{2 (4 \beta_1^2 - \beta_1^2 \beta_2 - 3 \beta_2)} \left[ -3 \beta_2 (1 + \beta_1) + \sqrt{3 \beta_2 \left[ 16 \beta_1^2 - \beta_2 (3 - \beta_1)^2 \right]} \right]                                                     \\       \nonumber
&& v = \xi_1 \xi_2 = -1 +  \frac{1 + \beta_1}{2 (4 \beta_1^2 - \beta_1^2 \beta_2 - 3 \beta_2)} \left[ -3 \beta_2 (1 + \beta_1) + \sqrt{3 \beta_2 \left[ 16 \beta_1^2 - \beta_2 (3 - \beta_1)^2 \right]} \right]. 
\end{eqnarray}
Thus, $\xi_1$ and $\xi_2$ for $\rho_{III}$ become
\begin{equation}
\label{negat-23}
\xi_1 = \frac{u + \sqrt{u^2 - 4 v}}{2}     \hspace{1.0cm}  \xi_2 = \frac{u - \sqrt{u^2 - 4 v}}{2}.
\end{equation}
Then, the R\'{e}nyi and von Neumann entropies of $\rho_{III}$ can be explicitly computed by inserting $\xi_1$ and $\xi_2$ in Eq. (\ref{negat-23}) 
into Eq. (\ref{m-entropy-1}) and (\ref{m-entropy-2}). 

Inserting Eq. (\ref{new5-1}) into Eq. (\ref{negat-22}), one can show straightforwardly 
\begin{equation}
\label{new5-4}
u = \frac{A_-}{x} \left[ 1 - \sqrt{1 - \frac{2 x}{A_+ + \sqrt{\tilde{A}}}} \right]                  \hspace{1.0cm}
v = -1 + \frac{A_+ + \sqrt{\tilde{A}}}{x} \left[ 1 - \sqrt{1 - \frac{2 x}{A_+ + \sqrt{\tilde{A}}}} \right].
\end{equation}
If $c_1 c_2 = |f|^2$, it is easy to show $v = 0$. Thus, Eq. (\ref{negat-23}) indicates that either $\xi_1$ or $\xi_2$ is zero depending on sign of $u$
when $c_1 c_2 = |f|^2$. 
For example, if $u$ is  positive,  $\xi_2 = 0$ and $\xi_1 = A_- / (A_+ + \sqrt{\tilde{A}})$. If, in this case, $c_1 = c_2 = f_R \equiv c$ with 
$\mbox{Im} f = 0$, $\xi_1$ reduces to 
\begin{equation}
\label{new5-5}
\xi_1 = \frac{c (a_1 + a_2 + 2 b)}{(a_1 a_2 - b^2) + \sqrt{(a_1 a_2 - b^2)^2 - c^2 (a_1 + a_2 + 2 b)^2}},
\end{equation}
which is exactly the same with $\xi_1$ of type I when $c = f$.

Above discussion implies that when $c_1 c_2 = |f|^2$, there may exist a tripartite pure state $\sigma_{ABC} = \psi (x_1', x_2', x_3') \psi^* (x_1, x_2, x_3)$,
which satisfies $\rho_{III} = \mbox{tr} \sigma_{ABC}$. In fact, one can construct two parameter ($\theta$, $\bar{x}$) family of 
$\psi(x_1, x_2, x_3)$, whose explicit form is 
\begin{eqnarray}
\label{puri5-1}
&&\psi(x_1, x_2, x_3: \theta, \bar{x})                               \\     \nonumber
&&= \sqrt{{\cal N}} 
\exp 
\left[ -r_1 e^{-i \phi_1} x_1^2 - r_2 e^{-i \phi_2} x_2^2 - \frac{\bar{x}}{2} x_3^2 - 2 r_3 e^{-i \phi_3} x_1 x_2 - 2 \sqrt{c_1 \bar{x}} e^{-i \theta}x_1 x_3-2 \sqrt{c_2 \bar{x}} e^{-i (\theta + \theta_f)} x_2 x_3   \right],
\end{eqnarray}
where $\theta_f = \mbox{arg} f$ and 
\begin{eqnarray}
\label{puri5-2}
&&{\cal N} = \frac{ \sqrt{\bar{x}} \sqrt{4 (a_1 - c_1) (a_2 - c_2) - (2 b + f + f^*)^2}}{\pi^{3 / 2}}     \hspace{1.0cm}
r_1^2 = (a_1 - c_1)^2 + 4 a_1 c_1 \cos^2 \theta                                                                        \nonumber    \\
&& r_2^2 = (a_2 - c_2)^2 + 4 a_2 c_2 \cos^2 (\theta + \theta_f)                                                   \hspace{1.0cm}
r_3^2 = c_1 c_2 + b^2 - 2 b \sqrt{c_1 c_2} \cos (2 \theta + \theta_f)                                                \\    \nonumber
&&\tan \phi_1 = \frac{c_1 \sin 2 \theta}{c_1 \cos 2 \theta + a_1}      \hspace{.5cm}
\tan \phi_2 = \frac{c_2 \sin (2 \theta + 2 \theta_f)}{c_2 \cos (2 \theta + 2 \theta_f) + a_2}     \hspace{.5cm}
\tan \phi_3 = \frac{\sqrt{c_1 c_2} \sin(2 \theta + \theta_f)}{\sqrt{c_1 c_2} \cos (2 \theta + \theta_f) - b}.
\end{eqnarray}
If we choose 
\begin{equation}
\label{puri5-3}
\bar{x} = \frac{1}{4 (a_1 - c_1) (a_2 - c_2) - (2 b + f + f^*)^2}    \hspace{1.0cm} \theta = 0,
\end{equation}
the parameters in Eq. (\ref{puri5-2}) reduce to
\begin{eqnarray}
\label{puri5-4}
&&{\cal N} = \frac{1} {\pi^{3/2}}   \hspace{.5cm} r_1 = a_1 + c_1  \hspace{.5cm}
r_2 = \sqrt{(a_2 - c_2)^2 + 4 a_2 c_2 \cos^2 \theta_f}                                                                           \\           \nonumber
&&r_3 = \sqrt{b^2 + c_1 c_2 - 2 b \sqrt{c_1 c_2} \cos \theta_f}          \hspace{1.0cm}    \phi_1 = 0         \\          \nonumber
&&\phi_2 = \tan^{-1} \left[ \frac{c_2 \sin 2 \theta_f}{c_2 \cos 2 \theta_f + a_2} \right]                     \hspace{.5cm}
\phi_3 = \tan^{-1} \left[ \frac{\sqrt{c_1 c_2} \sin \theta_f}{\sqrt{c_2 c_2} \cos \theta_f - b} \right].
\end{eqnarray}
If $c_1 = c_2 \equiv c$ and $\theta_f = 0$, it is easy to show that the pure state in Eq. (\ref{puri5-1}) exactly coincides with Eq. (\ref{puri3-2}).

\subsection{Type IV Bipartite State}
In this subsection we examine the following bipartite Gaussian state:
\begin{eqnarray}
\label{type4-1}
&&\rho_{IV} [x_1', x_2',: x_1, x_2] = A \exp \Bigg[ -a_1 (x_1^2 + x_2'^2) - a_2 (x_1'^2 + x_2^2)                           \\    \nonumber
&&\hspace{2.0cm}+ 2 b (x_1' x_2' + x_1 x_2) + 2 c (x_1 x_1' + x_2 x_2') + 2 f_1 x_1 x_2' + 2 f_2 x_1' x_2   \Bigg]
\end{eqnarray}
where $A = \sqrt{(a_1 + a_2 - 2 c)^2 - (2 b + f_1 + f_2)^2} / \pi$. The density matrix $\rho_{IV}$ is obtained from $\rho_{II}$ by taking a partial 
transposition ($x_1 \leftrightarrow x_1'$). In Ref. \cite{park19} it is shown that solving the eigenvalue equation (\ref{m-eigenvalue-1}) (with $j$ = IV) is 
highly difficult because we cannot find suitable change of variables to factorize Eq. (\ref{m-eigenvalue-1})  into the two single-party eigenvalue equations.
In fact, we do not know whether the factorization is possible or not.
Thus, we will compute the R\'{e}nyi and von Neumann entropies of $\rho_{IV}$ without solving the eigenvalue equation explicitly as we did in the previous section. Here, we assume that the eigenvalues of
$\rho_{IV}$ is also represented as Eq. (\ref{m-eigenvalue-2}). 

First, we note
\begin{equation}
\label{new6-1}
\tilde{\beta}_1 \equiv \mbox{tr} \rho_{IV}^2 = \sqrt{\frac{Y_1}{Y_2}}                \hspace{1.0cm}
\tilde{\beta}_2 \equiv \mbox{tr} \rho_{IV}^3 = \frac{4 Y_1}{Y_1 + 3 Y_2 - 12 y}
\end{equation}
where
\begin{equation}
\label{new6-2}
Y_1 = (a_1 + a_2 - 2 c)^2 - (f_1 + f_2 + 2 b)^2            \hspace{.5cm}
Y_2 = (a_1 + a_2 + 2 c)^2 - (f_1 + f_2 - 2 b)^2       \hspace{.5cm} y = c^2 - f_1 f_2.
\end{equation}
Let $Y_2 \pm Y_1 = 2 B_{\pm}$ and $\tilde{B} = Y_1 Y_2$, where $B_+ = (a_1 + a_2)^2 + 4 c^2 - (f_1 + f_2)^2 - 4 b^2$ and 
$B_- = 4 c (a_1 + a_2) +  4 b (f_1 + f_2)$. Then, if one follows the procedure of previous subsection, $\xi_1$ and $\xi_2$ of $\rho_{IV}$ are given by 
\begin{equation}
\label{new6-3}
\xi_1 = \frac{\bar{u} + \sqrt{\bar{u}^2 - 4 \bar{v}}}{2}     \hspace{1.0cm}
\xi_2 = \frac{\bar{u} - \sqrt{\bar{u}^2 - 4 \bar{v}}}{2}
\end{equation}
with
\begin{equation}
\label{new6-4}
\bar{u} = \frac{B_-}{4 y} \left[ 1 - \sqrt{1 - \frac{8 y}{B_+ +  \sqrt{\tilde{B}}}} \right]                 \hspace{.5cm}
\bar{v} = -1 + \frac{B_+ + \sqrt{\tilde{B}}}{4 y}  \left[ 1 - \sqrt{1 - \frac{8 y}{B_+ +  \sqrt{\tilde{B}}}} \right].
\end{equation}   
If $c^2 = f_1 f_2$, $\bar{u}$ and $\bar{v}$ become
\begin{equation}
\label{new6-5}
\bar{u} = \frac{B_-}{B_+ + \sqrt{\tilde{B}}}    \hspace{1.0cm}   \bar{v} = 0.
\end{equation}   
Thus, in this case, either $\xi_1$ or $\xi_2$ is zero depending on the sign of $\bar{u}$. In spite of this fact we cannot find any 
tripartite pure state $\sigma_{ABC}$ whose substate coincides with $\rho_{IV}$. This seems to be due to the fact that $\rho_{IV}$ cannot 
arise in the usual zero-temperature harmonic oscillator system like $\rho_{II}$.

\section{Conclusions}
In this paper we derive the R\'{e}nyi and von Neumann entropies for four types of the bipartite Gaussian states. For the case of type I and type II we solve 
the eigenvalue equation (\ref{m-eigenvalue-1}) (with $j$ = I, II) analytically by factorizing it into two single-party eigenvalue equations. For the case of 
type III we use a different method. Since the derivation of R\'{e}nyi and von Neumann entropies does not need  the normalized eigenfunctions, we 
derive the eigenvalues of Eq. (\ref{m-eigenvalue-1}) (with $j$ = III) without deriving the eigenfunction explicitly. This method is valid because of the fact that 
 the eigenvalue of $\rho_{III}$ is represented as Eq. (\ref{m-eigenvalue-2}) as shown in Ref. \cite{park18-2}. For the case of type IV we derive
 the entropies under the assumption that the eigenvalue of Eq. (\ref{m-eigenvalue-1}) (with $j$ = IV) is also represented as Eq. (\ref{m-eigenvalue-2}).

Of course, there might be many bipartite Gaussian states, which do not belong to these four types. We do not know whether the R\'{e}nyi and 
von Neumann entropies of all bipartite Gaussian states can be exactly derived or not. Furthermore, we do not know how to solve the eigenvalue 
equation (\ref{m-eigenvalue-1}) for non-Gaussian state. For example, let us consider the general Gaussian or non-Gaussian bipartite 
state $\rho [x_1', x_2': x_1, x_2]$. Then, its eigenvalues $\lambda_{mn}$ satisfy 
\begin{equation}
\label{determinant}
\mbox{det} \bigg[ \rho [x_1', x_2': x_1, x_2] - \lambda_{mn} \delta (x_1 - x_1') \delta (x_2 - x_2') \bigg] = 0.
\end{equation}
If the determinant of Eq. (\ref{determinant}) can be computed, it is possible to derive the R\'{e}nyi and von Neumann entropies for general state $\rho$ without deriving the 
normalized eigenfunctions. Probably, the computation of the determinant may need to adopt some regularization scheme to escape the 
infinity arising due to $\delta$-functions. We hope to explore this issue in the future.


\end{document}